\documentstyle[twoside,11pt]{article}
\newcommand{\be}{\begin{equation}}
\newcommand{\ee}{\end{equation}}
\newcommand{\g}{\Gamma (\bar{\nu})}
\newcommand{\quadcap}{\left( \frac{k}{2H} \right)^{-2\bar{\nu}}}

\newcommand{\s}{\sigma}
\begin{document}
\baselineskip=18pt
\setcounter{page}{1}

\title{ \bf{Quantum Fields in an Expanding Universe}}

\author{F. Finelli\thanks{e--mail: finelli@tesre.bo.cnr.it} $^{\, 1,2}$, 
A. Gruppuso\thanks{e--mail: gruppuso@bo.infn.it} $^{\, 2}$, and G.
Venturi\thanks{e--mail: armitage@bo.infn.it} $^{\, 2}$ \vspace{.5cm} \\
{\small\it $^1$ Istituto Te.S.R.E./CNR, via Gobetti 101, 40129 Bologna
Italy}\\
{\small\it $^2$ Dipartimento di Fisica, Universit\`a degli Studi di
Bologna
and INFN} \\
{\small\it via Irnerio 46, 40126 Bologna Italy}}
%
\pagestyle{plain}
\maketitle
\begin{abstract}
We extend our analysis for scalar fields in a 
Robertson-Walker 
metric to the electromagnetic field
and Dirac fields by the method of invariants. 
The issue of the relation between conformal properties 
and particle production is re-examined and it is verified that the
electromagnetic and massless 
spinor actions are conformal invariant, while the massless conformally 
coupled scalar field  is not.
For the scalar field case it is pointed out that the violation 
of conformal simmetry due to surface terms, although ininfluential for the 
equations of motion, does lead to effects in the quantized theory.

\end{abstract}

\section{Introduction}


The interest in quantum field theory (henceforth QFT) 
in classical curved space-times has 
never ceased especially after the development of particle production
in cosmological 
spacetimes, initiated by L. Parker in 1968 \cite{parker}, and 
the discovery of black hole 
evaporation by S. W. Hawking in 1974 \cite{hawking}.
Such a semiclassical treatment for gravity should hold in a region between 
the Planck length and the Compton wavelength (and 
certainly greater than this) of the matter field considered 
and could be a useful route towards  
understanding the principal ingredients for a theory 
which unifies gravity and quantum mechanics.  
All the theoretical motivations for studying QFT 
in curved spacetimes are 
corroborated by a possible application of the results to cosmology: 
a spectrum of quantum fluctuations during inflation 
could be related to the primordial spectrum 
of cosmological perturbations \cite{k&tlinde}. 

For the cosmological case particle production is related to the simple 
and well studied case of a harmonic oscillator with time dependent 
parameters. The time dependence of the metric leads to the appearance 
of different vacua for differing times during the evolution of a free field.
When exact solutions to the field equations of motion are not 
known, a systematic treatment based on the adiabatic approximation 
can be found in the literature \cite{bd}. For this reason it 
is a usual practice to 
rescale fields by a time dependent factor 
so as to obtain suitable expressions in order to 
to implement an adiabatic expansion around a time independent  
solution. This rescaling factor 
is also related to the conformal weight of the field which leaves the 
equations of motion invariant under a conformal trasformation in both 
the metric and the field. Since the Robertson-Walker
(henceforth RW) metric is conformally related 
to a Minkowski metric one has a correspondence between fields in 
a RW metric and in flat spacetimes with time dependent masses. 
For the case of massless conformally coupled fields the time dependent 
parameters in the equations of motion disappear 
and the theory in such a curved spacetime corresponds to a theory 
in Minkowski spacetime.

The method of invariants \cite{lewis} allows one to exactly 
quantize a harmonic oscillator with time dependent coefficients. 
The application of this method in QFT 
improves on the adiabatic approximation and 
allows one to introduce a vacuum and a Fock space associated with the 
quantum invariants.
The use of the method of quantum invariants \cite{lewis} has been 
previously applied, within the context of cosmology, to a quantized 
scalar field in a de Sitter space-time with a flat spatial section 
\cite{gao0, desitter}. Such an approach also arises naturally within a 
Born-Oppenheimer context, for the matter-gravity 
system \cite{bv}, in a simple minisuperspace model when the semiclassical 
limit is taken for gravity and fluctuations are neglected \cite{born}.

In the previously studied scalar case a massive scalar field with 
a non-minimal coupling was examined \cite{desitter}.
It was found that the expectation value of the Hamiltonian of a Fourier 
mode of the field grows
with time even in the massless conformally coupled case. However, 
for a non-minimally coupled scalar field 
the Hamiltonian density differs from the $00$ component of the 
energy-momentum tensor, and the latter vanishes for zero mass and
conformal coupling.
The usual statement of no particle production in the massless 
conformally coupled case should then be applied directly to the Einstein
equations: only the conformal anomaly is a source for the Einstein tensor 
in this case. Further in \cite{desitter} it was also pointed out 
that the usual conformal rescaling does not leave the 
action invariant because of the presence of a boundary term.

It is then natural to re-examine, within the context of quantum invariants, 
the results previously obtained for the electromagnetic field and spin 
$1/2$ particles. This shall be done in the next two sections respectively.
In section 4 we discuss the conformal properties of the action considered, 
while in section 5 
we analyze in detail the problems arising from the quantization 
of rescaled and non-rescaled fields. 
Finally in section 6 we present our conclusions.

\section{Electromagnetic Field}

In this and the following sections we consider a RW line element
\be
ds^2 = -d\tau^2 +a^2 (\tau) g_{ij} dx^i dx^j
\label{rwmetric}
\ee
where $g_{ij}$ is the three metric for a flat three-space.
In the absence of charges the Lagrangian density for the electromagnetic 
potential $A_\mu$ is given by:
\begin{eqnarray}
{\cal L}_{EM} &=& \sqrt{-g} \left[ -\frac{1}{4} F_{\mu \nu} F^{\mu \nu} 
\right]  \nonumber \\ &=& \frac{a^3}{4} \left[ + \frac{2}{a^2} F_{0j}^2 
- \frac{1}{a^4} F_{ij}^2 \right]
\label{emlag}
\end{eqnarray}
where $F_{\mu \nu}\equiv \nabla_\mu A_\nu -\nabla_\nu A_\mu = 
\partial _\mu A_\nu -\partial_\nu A_\mu $ (with $\nabla _\mu$ the covariant 
derivative) because of the simmetry in the lower indices
of the Cristoffel connections.
On choosing the generalized Lorentz gauge $ \nabla _\mu A^{\mu} =0$ and 
$ A_{0}=0$  (which is always possible in the source-free case) 
and considering a RW metric, we obtain the so called radiation gauge 
($\vec \nabla \cdot \vec A =0 $). On then substituting 
$A_\mu = (A_0, \vec{A})$ in Eq. (\ref{emlag}) and using the radiation gauge 
one has 
\begin{eqnarray}
F_{0j} &=& \dot A_j \nonumber \\
F_{ij} &=&  (\vec{\nabla} {\times} \vec{A})_{ij} \,,
\end{eqnarray}
where by the dot we denote a derivative with respect to the proper time
$\tau$. Let us 
note that the vector potential $\vec{A}$ has been chosen to be the 
covariant $A_{i}$, 
and this is a {\em natural} choice if we wish to identify $F_{\mu \nu}$
with a two form (metric independent). Further the choice of $\vec{A}$ 
as a generalized coordinate with its subsequent quantization allows us 
to use the method of invariants for all the Fourier components of 
$\vec{A}$ \cite{gao} (this would not be possible for the case of the 
identification with the controvariant form $A^{\mu}$).
The final expression one then obtains for the
electromagnetic 
Lagrangian density is 
\be
{\cal L}_{EM} = \frac{1}{2} \left\{ a \dot{\vec{A}}^2 + \frac{1}{a} 
\vec{A} \cdot \nabla^2 \vec{A} - \frac{1}{a} \vec{\nabla} \cdot
\left[ A_i  \vec{\nabla} A_i - (\vec{A}\cdot \vec{\nabla})
 \vec{A} \right] \right\} \,.
\ee

We expand $\vec{A}$ as 
\be
\vec{A} = \frac{1}{\sqrt{V}} \sum_{\vec{k}\,,\lambda} 
\left[ c_k^{(\lambda)} (\tau) \vec{\varepsilon}^{\,(\lambda)} 
e^{i \vec{k} \cdot \vec{x}} 
+ c_k^{(\lambda) *} (t) 
\vec{\varepsilon}^{(\lambda) } 
e^{-i \vec{k} \cdot \vec{x}} \right] \,,
\ee
where $\lambda$ runs over the (two) polarization states and 
$\vec{\varepsilon}^{\,(\lambda)}$ is a unit vector which satisfies  
$\vec{k} \cdot \vec{\varepsilon}^{\,(\lambda)}=0$ and 
$\vec{\varepsilon}^{\,(\lambda)} \cdot
\vec{\varepsilon}^{\,(\lambda')}=\delta_{\lambda\,\lambda'}$.
On further separating $c_k^{(\lambda)}$ into real and 
imaginary parts 
\be
c_k^{(\lambda)} = \frac{1}{\sqrt{2}} \left( c_{k\,1}^{(\lambda)}
+ ic_{k\,2}^{(\lambda)} \right)
\ee
the action becomes:
\be
S = \sum_{k, i, \lambda} S^{(\lambda)}_{k \, i} = 
\frac{1}{2}\sum_{k, i, \lambda} \int \left[ a \dot c_{k\,i}^{(\lambda)\,2}
- \frac{k^2}{a} c_{k\,i}^{(\lambda)\,2} \right] d\tau \,.
\ee
Thus we see that the different modes $k, i, (\lambda)$ decouple and 
one then obtains the following Hamiltonian for each mode:
\be
H_{i\,k}^{(\lambda)} = \frac{1}{2}
\left( \frac{\pi_{i,k}^{(\lambda)\,2}}{a} + a
\omega_k^2 c_{k\,i}^{(\lambda)\,2} \right) \,,
\label{emham}
\ee
where $\pi_{i,k}^{(\lambda)}= a \dot c_{k\,i}^{(\lambda)}$ and 
$\omega_k^2 =k^2/a^2$. The classical equation of motion is:
\be
\ddot{c}_{k\,i}^{(\lambda)} + \frac{\dot a}{a}\,
\dot{c}_{k\,i}^{(\lambda)} +
\omega_k^2 c_{k\,i}^{(\lambda)}= 0
\,
\label{emclass}
\ee
and we proceed in analogy with the scalar field case \cite{desitter}.
On canonically quantizing, the Hamiltonian 
in Eq. (\ref{emham}) can
be factorized as (henceforth we shall denote collectively 
$\bf{k}, i, (\lambda)$ by $\sigma$ and retain the subscript $k$ 
only when relevant):
\be
\hat{H}_{\s} = \hbar \omega_k \left(\,
\hat a_{\s}^{\dagger} \,
\hat a_{\s}
+ \frac{1}{2}\,\right)
\ee 
with 
\be
\begin{array}{c}
\hat{a}_{\s} 
= \left( \frac{a \omega_k}{2 \hbar}
\right)^{\frac{1}{2}}
\left( \hat c_{\s} + i\,{\hat \pi_{\s} \over a\omega_k} \right) \,,
\\
\\
\hat{a}_{\s}^{\dagger} = \left( \frac{a \omega_k}{2 \hbar}
\right)^{\frac{1}{2}}
\left( \hat c_{\s} - i\,{\hat \pi_{\s} \over a\omega_k} \right)
\ 
\end{array}
\ee
and $[\hat{a}_{\s}, \hat{a}_{\s}^{\dagger}]
=\delta_{\s\, \s'}$.

As we mentioned, a suitable method for the study of time dependent quantum 
systems is that of invariants \cite{lewis}.
In particular a  hermitian operator $\hat I$ which satisfies:
\be
{\partial \hat I_{\s} (\tau) \over \partial\,\tau}
- \frac{i}{\hbar} [\hat I_{\s} (\tau), \hat H_{\s} (\tau)]=0 \,
\ee
is an invariant.
The invariant $\hat I_{\s}$ has real, time independent,
eigenvalues
and
in our case, can be decomposed in terms of basic linear invariants
\cite{gao}:
\be
\begin{array}{c}
\hat I_{b\, \s}(\tau) \equiv
e^{i\Theta_k(\tau)} \hat{b}_{\s} (\tau) \equiv { e^{i\Theta_k} \over
\sqrt{2\,\hbar}}\,
\left[{\hat c_{\s} \over \rho_k}+
i\,\left(\rho_k\,\hat \pi_{\s} - a \dot{\rho}_k \hat c_{\s} \right)\right] \,,
\\
\\
\hat I_{b\,\s}^{\dagger} (\tau) \equiv
e^{-i\Theta_k (\tau)} \hat{b}_\s(\tau) ^{\dagger} \equiv { e^{-i\Theta_k} \over
\sqrt{2\,\hbar}}\,\left[{\hat c_\s \over \rho_k}-
i\,\left(\rho_k\,\hat \pi_\s - a \dot{\rho}_k \hat c_\s
\right) \right]
\,, \label{inv}
\end{array}
\ee
where $\rho_k(\tau)$ is real and satisfies \cite{lewis}:
\be
\ddot{\rho}_k + \frac{\dot a}{a} \dot{\rho}_k
+ \omega^2_k \rho_k = \frac{1}{a^2 \rho_k^3}
\label{rho}
\ee
with:
\be
\Theta_k(\tau)=\int_{- \infty}^t
{d\tau'\over a(\tau')\,\rho_k^2(\tau')} \,.
\ee
and:
\be
[\hat{b}_\s, \hat{b}_{\s}^{\dagger}]
= \delta_{\s \s'} \, .
\ee
Let us now note that the above equations (\ref{inv}) may be rewritten in terms 
of the classical solutions $c_\sigma$ to Eq. (\ref{emclass}) through 
$c=\rho_k e^{-i \Theta_k}$: 
\be
\hat I_{b \, \sigma} (\tau) = i (c^*_\sigma \hat \pi_\sigma - 
\pi_{\sigma}^* \hat c_\sigma )
\ee
and the quadratic, hermitian, adiabatic invariant originally
introduced in \cite{lewis} is given by:
\be
\hat I_{\s} (\tau) =
\hbar \left(\hat b_\s^{\dagger} \, \hat b_\s + \frac{1}{2} \right) 
= \frac{1}{2}
\left[ \frac{\hat c_\s^{2}}{\rho_k^2} + (\rho_k \hat \pi_\s - a
\dot{\rho}_k \hat c_\s)^2 \right]
\ ,
\ee

The linearly independent solutions to the equation of motion
(\ref{emclass}) are the Bessel functions: 
\be
c = \eta^{1/2} \left\{ \begin{array}{c}
J_{1/2} (k\eta) \\
N_{1/2} (k\eta)
\end{array} \right. 
\ee
where $\eta$ is
the conformal time and the above solutions are true both for 
de Sitter $a=-\frac{1}{H\eta}$ (H is the Hubble constant, $-\infty<\eta<0$)
and for power behaviour $a=\eta^p$ (with $p$ a positive real number). 
The general solution to Eq. (\ref{rho}) 
can be written as a non-linear combination of the 
solutions to the equations of motion as shown in \cite{desitter}:
\be
\rho= \eta^{1/2} \left[ A J_{1/2}(k\eta) + B N_{1/2}(k\eta)
 + 2(AB-\frac{\pi^2}{4})
J_{1/2}(k\eta) N_{1/2}(k\eta) \right]^{\frac{1}{2}}
\ee
where $A, B$ are real constants ($k$ independent because 
of the spatial simmetry of RW spacetime). 
On choosing $A=B=\pi/2$ one has 
\be
\rho = \frac{1}{\sqrt k}
\label{emfinal}
\ee
which is independent of $\eta$. The choice $A=B$ is associated with 
the adiabatic vacuum at early times 
or the adiabatic vacuum for wavelengths $2 a \pi/k$ which 
are well inside the Hubble radius $a/ \dot a = H^{-1} $ 
(see \cite{desitter}), i.e. the Bunch-Davies vacuum. 

>From Eq. (\ref{emfinal}) we see that the annihilation operators 
$\hat a$ and $\hat b$ coincide for all times which implies 
\be
\hat{H}_{k\,i}^{(\lambda)} = \omega_k \hat{I}_{k\,i}^{(\lambda)} \,.
\ee
Hence for a (massless) photon 
the initial adiabatic vacuum remains such for all times, leading 
to a null photon production in RW spacetime, and its energy is redshifted
as expected for radiation. 
We end by observing that if one relaxes the assumption of an adiabatic vacuum 
at early times, i.e. allowes $A \ne B$, one has a  
number of photons oscillating around the initial number, without 
a net growth.

\section{Dirac Spinor Field}

For the case of a massive spin $\frac{1}{2}$ field $\Psi$ the 
lagrangian density is:
\be
{\cal L} = -\sqrt{-g} \left\{ \frac{i}{2} \left[ \bar{\Psi} \gamma^\mu 
(\nabla_\mu \Psi) - (\nabla_\mu \bar{\Psi} )\gamma^\mu \Psi \right] 
+ \mu \bar{\Psi} \Psi \right\}  
\ee
which in RW spacetimes can be rewritten as 
\begin{eqnarray}
{\cal L} &=& a^3 \left[ \frac{i}{2} (\Psi^{\dagger} \dot \Psi - 
\dot{\Psi}^{\dagger} \dot \Psi)- \Psi^\dagger \gamma_4 ({\vec{\gamma}\over{a}}
 \cdot \vec{\nabla} + \mu) \Psi \right] \nonumber \\
&\equiv& a^3 \left[ \frac{i}{2} (\Psi^{\dagger} \dot \Psi - 
\dot{\Psi}^{\dagger} \dot \Psi)- \Psi^\dagger M \Psi \right]
\end{eqnarray}
where $\mu$ is the inverse Compton wavelength of the spinor field and 
we shall use the Pauli-Dirac representation for the $\gamma$ matrices. 
On expanding 
\be
\Psi = \frac{1}{\sqrt{V}} \sum_{\bf k} 
\Psi_{k} e^{i {\vec k} \cdot {\vec x}} \label{EFourier}
\ee
one obtains an action:
\begin{eqnarray}
S &=& \sum_k S_k = \frac{1}{\sqrt{V}} \sum_k 
\int d \tau a^3 \left[ \frac{i}{2} (\Psi^\dagger_k \dot \Psi_k - 
\dot \Psi_k^\dagger \Psi_k) - \Psi_k^\dagger (\frac{i}{a} \gamma_4 
\vec{\gamma} \cdot \vec{k} + \gamma_4 \mu) \Psi \right]
\nonumber  \\
&\equiv& \frac{1}{\sqrt{V}} \sum_k 
\int d \tau a^3 \left[ \frac{i}{2} (\Psi^\dagger_k \dot \Psi_k - 
\dot \Psi_k^\dagger \Psi_k) - \Psi_k^\dagger M_k \Psi_k \right]
\end{eqnarray}
and again one can consider each Fourier mode separately and obtain 
a single mode Hamitonian:
\be 
H_k = \psi_k^\dagger M_k \psi_k
\ee
where $\psi_k = a^{3/2} \Psi _k$ and the matrix $M_k$ is: 
\be
M_k = \left( \begin{array}{cc}
\mu & \vec{\sigma} \cdot \frac{\vec{k}}{a} \\
\vec{\sigma} \cdot \frac{\vec{k}}{a}  & -\mu 
\end{array} \right) \,.
\label{fermmat}
\ee

>From the Lagrangian density and eq.(\ref{EFourier}) one obtains a
classical equation of motion:
\be
\left[ - i \frac{\partial}{\partial t} + i \gamma_4 \vec{\gamma}\cdot 
\frac{\vec{k}}{a} + \gamma_4 \mu \right] w_k^{(r)} = 0
\ee
which has been previously solved for a de Sitter space-time \cite{barut} 
obtaining:
\be
w_{1\,,\bf k}^{(r)} = \frac{1}{a^{1/2}} \left(
\begin{array}{c}
i Z_{\nu} \chi ^{(r)}\\
\frac{\vec \sigma \cdot \vec k}{k} Z_{\nu -1} \chi ^{(r)}
\end{array} \right) \mbox{ r=1,2}
\label{uno}
\ee
and
\be
w_{2\,,\bf k}^{(r)} = \frac{1}{a^{1/2}} \left(
\begin{array}{c}
- \frac{\vec\sigma \cdot \vec k}{k} Z_{\nu^* -1} \chi ^{(r)}\\
-i Z_{\nu^*} \chi ^{(r)}
\end{array} \right) \mbox{ r=3,4}
\label{due}
\ee
where $r=1,3 (2,4)$ correspond to spin up (down) for the Pauli spinors 
$\chi ^{(r)}$
and $Z_\nu (k |\eta|)$ are Bessel functions ($J_\nu$, $N_\nu$ or a
combination of them). Further $\nu=\frac{1}{2} - i\frac{\mu}{H}$ 
and we introduce a normalization factor $N_{a \, k}$ determined by 
\be
N_{a \, k}^2 w^{(r)\dagger}_{k} w^{(r')}_{k} = \delta_{r\, r'}
\ee

It is now of interest to examine the adiabatic limit for which the 
solutions (\ref{uno}) and (\ref{due}) reduce to the 
usual static solutions:
\be
u_{\bf k}^{(r)} = \left(
\begin{array}{c}
\chi ^{(r)}\\
\frac{\vec \sigma \cdot \vec k}{a(\omega_k+\mu)} \chi ^{(r)}
\end{array} \right) e^{-i \omega t}  \quad r=1,2
\label{unobis}
\ee
and
\be
u_{\bf k}^{(r)} = \left(
\begin{array}{c}
- \frac{\vec \sigma \cdot \vec k}{a(\omega_k+\mu)} \chi ^{(r)}\\
 \chi^{(r)}
\end{array} \right) e^{i \omega t}  \quad r=3,4
\label{duebis}  
\ee
respectively, where $\omega=(\frac{k^2}{a^2} + \mu^2)^{1/2}$
and $r=1,3 (2,4)$ again refers to spin up (down) for the Pauli spinors.
As in the previous case we introduce a normalization factor $N_k$ determined by:
\be
 N_k^2 u_{\bf k}^{(r) \dagger} u_{\bf k}^{(r')}= \delta_{r\, r'} \, .
\ee 
We expect that the reduction to the usual static solutions occurs for very
early times ($\tau \rightarrow - \infty$)
or for wavelengths which are very small compared to the 
de Sitter horizon $H^{-1}$. In such a limit ($-k\eta=\frac{k}{H} e^{-Ht} 
\rightarrow \infty$) Eq. (\ref{unobis}) leads to 
\be
u_{\bf k}^{(r)} \rightarrow \frac{1}{\sqrt{2}} \left( 
\begin{array}{c}
-\chi^{(r)} \\
\frac{\vec \sigma \cdot \vec k}{k} \chi^{(r)} 
\end{array}  \right) e^{i k \eta}  \quad r=1,2
\ee 
which requires that one take the Hankel functions:
\be
Z_\nu = H_\nu^{(1)}
\ee
and a similar result only involving $H^{(2)}_{\nu ^*}$ is obtained from Eqs. 
(\ref{due}) and (\ref{duebis}). Again, as in previous cases, we have a two 
parameter
set of solutions: it is only by requiring agreement for a particular 
(early time) limit with the adiabatic solutions that we constrain them.  

We may quantize the system by postulating the usual canonical 
anticommutation relations 
\be
\left\{ \hat \psi^\dagger_{\bf k},\hat \psi_{\bf k} \right\} = \hbar 
\ee
with all the other anticommutators being zero. 
If we employ the adiabatic (static) solutions Eqs. (\ref{unobis}) and 
(\ref{duebis}) one may expand  
\be
\hat \psi = \frac{1}{\sqrt V} \sum_{ k} e^{i {\vec k} \cdot {\vec
x}} \sum_r N_k u^{(r)}_{\bf k} \hat a_{\bf k}^{ (r)} 
\ee   
with $\left\{\hat a^\dagger_{\bf k} , \hat a_k
\right\} = \hbar$  and on
substituting in the quantum Hamiltonian we obtain: 
\be  
\hat H_{\bf k} =\hat {\psi}_k^\dagger M_k \hat {\psi}_k =
 \sum_{r=1,2} \omega_k \hat a^{(r)\,\dagger}_{\bf k} 
\hat a^{(r)}_{\bf k} - \sum_{r=3,4} \omega_k \hat
a^{(r)\,\dagger}_{\bf k}
\hat a^{(r)}_{\bf k}
\ee
On then introducing a vacuum consisting of a "Dirac sea" of negative energy
states one has the usual interpretation of the destruction 
of a negative energy as the creation of a positive energy antiparticle. 
Thus through the usual replacements
\begin{eqnarray}
\hat a_{\bf k}^{(s)} = \hat a_{\bf k}^{(r)} \; ; 
u_{\bf k}^{(s)} = u_{\bf k}^{(r)}  \; \mbox{with  r=s for $r=1,2$ }\nonumber \\
(-)^s c_{\bf k}^{(s)\,\dagger} = \hat a_{\bf k}^{(r)}
\; ;(-)^s v_{\bf k}^{(s)} = u_{-\bf k}^{(r)}\; 
\mbox{with $s=1 (2)$ for $r=4(3)$}
\label{hole}
\end{eqnarray}
one obtains 
\be
\hat \psi = \frac{1}{\sqrt V} \sum_{{\bf k}\,s} N_k\left( e^{i {\vec k} 
\cdot {\vec x}} \hat a_{\bf k}^{(s)} u_{\bf k}^{(s)} 
+ e^{- i {\vec k}
\cdot {\vec x}} \hat c_{\bf k}^{(s) \, \dagger} v_{\bf k}^{(s)} \right)
\ee

For this case, as in the previous section, one may also construct linear 
invariants $\hat I$ by using Heisenberg fields and classical solutions through:
\be
 \hat I_{\bf k}^{(r)} = w_{\bf k}^{(r)\, \dagger} \hat \psi_{\bf k}
 =N_{a \, k}^{-1}\hat b_{\bf k}^{(r)}
\ee
where $\left\{ \hat b_{\bf k}^{(r)\, \dagger} ,\hat b_{\bf k}^{(r)}
\right\} = \hbar $ 
which, on using the anticommutation relations for $\hat \psi $ and the 
classical equations of motions for $w$, satisfy: 
\be
{\partial \hat I_{\bf k}^{(r)} (t) \over \partial\,t}
- \frac{i}{\hbar} [\hat I_{\bf k}^{(r)} (t), \hat H_{\bf k} (t)]=0 \,.
\ee
Correspondingly one may expand
\be
\hat \psi  =\frac{1}{\sqrt V} \sum_{\bf k} e^{i {\vec k} \cdot {\vec x}} 
\sum_r N_{a \, k} w^{(r)}_{\bf k} \hat b_{\bf k}^{(r)}
\ee
and again one may introduce a vacuum consisting of a 'sea' of the equivalent 
of the negative energy states ($r=3,4$) with the usual interpretation of 
a destruction operator as the creation of a 'hole'. Thus the relations 
equivalent to eqs. (\ref{hole}) hold for the $b$ operators, where 
the $d$ operators and the functions $y$ are introduced for $r=3,4$ 
(corresponding to $c$ and $w$), 
obtaining 
\be
\hat \psi = \frac{1}{\sqrt V} \sum_{{\bf k}\,s}N_{a \, k}\left( e^{i {\vec k} 
\cdot {\vec x}} \hat b_{\bf k}^{(s)} w_{\bf k}^{(s)}  
+ e^{- i {\vec k}
\cdot {\vec x}} \hat d_{\bf k}^{(s) \, \dagger} y_{\bf k}^{(s)} \right)
\ee
It is clear that the creation and destruction operators in the 
two bases are related by a Bogoliubov transformation whose coefficients 
may be easily determined;
thus for example the creation 
of an invariant fermion quantum will correspond to a mixture 
of the creation of a Dirac particle and the destruction of a 
Dirac antiparticle. Further the quantum invariants allow 
us to define an invariant vacuum 
(the fermion equivalent of the Bunch-Davies vacuum \cite{bunchdavies}) by
\be
b_{\bf k}^{(s)} |0\rangle_b = d_{\bf k}^{(s)} |0\rangle_b = 0
\ee

We may now examine particle production from the vacuum during 
a de Sitter expansion. In order to do this it will be sufficient 
to consider the expectation value of the Hamiltonian with respect to the 
above vacuum. On using the $\psi$ expansion in the 
invariant basis it is straightforward to obtain 
\begin{eqnarray} 
\lim_{a \rightarrow \infty}  { }_b\langle 0|\hat H_{\bf k} (t) |0\rangle_b  
 \!\!\!\!\! &=& \!\!\!\!\!
\lim_{a \rightarrow \infty} N_{a k}^2 \sum_{r=3,4} w_k^{(r)\dagger}M_k w_k^{(r)}
 =\nonumber \\
\quad = 2\mu \left\{ \left| {\pi \left[ 1+i \cot {(\nu-1)\pi}\right] 
\over{ \Gamma (\nu)^2}} \right| ^2 -1 \right\} & & \!\!\!\!\!\!\!\!\!\!\!\!
\left\{ \left| { \pi \left[ 1+i \cot {(\nu-1)\pi} \right] 
\over{ \Gamma (\nu)^2}}\right| ^2 +1 \right\} ^{-1} 
\label{limferm}
\end{eqnarray}
which is constant, that is the number of fermions does not increase in
time, and in particular we note that the RHS of Eq. (\ref{limferm}) 
is zero for $\mu = 0$ (actually there is a non-leading term ${\cal O} (k/a)$
corresponding to radiation). 
Let us note that we have not concerned ourselves
with renormalizing the vacuum energy since we are just interested
in changes in it.

One may also consider the expectation value of the Hamiltonian 
with respect to other states, such as coherent states. Such states, for 
fermions, essentially consist of a superposition of zero and one fermion 
states, because of the Pauli exclusion principle. It is straightforward to 
verify that in this case also the expectation value of the Hamiltonian 
is asymptotically constant in time and is of order ${\cal O} (k/a)$ 
for $\mu = 0$. This of course means that in the de Sitter expansion 
the number of fermions in a given mode is asymptotically a constant.

In the next section we shall re-examine the results of this and the previous 
section by examining the properties of the action rather than the 
equations of motions under conformal transformations.

\section{Conformal Invariance}

Let us consider a general line element:
\be
ds^2 = g_{\mu \nu} dx^\mu dx^\nu
\ee
and the action of a massless conformally coupled scalar field:
\be
{S}_{\Phi} = - \int d^4x \sqrt{-g} \left[ 
\frac{1}{2}
g^{\mu \nu} \nabla_{\mu} \Phi \nabla_{\nu} \Phi 
+ \frac{1}{12} R {\Phi}^2 \right] \,.
\label{tilded}
\ee
We may now perform a conformal trasformation:
\begin{eqnarray}
g_{\mu \nu} \rightarrow \tilde{g}_{\mu \nu} = \Omega^2 g_{\mu \nu} 
\nonumber \\
\Phi \rightarrow \tilde \Phi = \Omega^{-1} \Phi
\label{conftras}
\end{eqnarray}
with $\Omega $ real, non-zero, continuous
and obtain for the change in the action 
\begin{eqnarray}
\tilde{S}_{\tilde {\Phi}} - S_{\Phi} &=& \frac{1}{2} \int_V d^4 x \sqrt{-g} 
\nabla^{\mu} \left( \Phi^2 \nabla_{\mu} \ln \Omega \right)
\nonumber \\ 
&=& \frac{1}{2} \int_{\Sigma(V)} d^3 \sigma \sqrt{-g} 
n^{\mu} \Phi^2 \nabla_{\mu} \ln \Omega 
\end{eqnarray}
where $n_\mu$ is a unit vector perpendicular to the three dimensional 
surface $\Sigma$ containing V. 

In particular for the RW metric (\ref{rwmetric}) the surface term becomes
\be
-\frac{1}{2} \int_V d^3 x d\tau
{\partial \over{\partial \tau}} \left( \Phi^2 \dot a \right)
= - \frac{1}{2} \int_V d\eta d^3 x
{\partial \over{\partial \eta}} \left(\Phi^2 {1\over a}
{\partial a \over{\partial \eta}} \right)
\ee
in agreement with our previous result \cite{desitter}. On the right
hand side we have introduced the conformal time $\eta$ with $a d\eta =
d\tau$ with which the metric (\ref{rwmetric}) becomes conformally flat:
\be
ds^2 = a^2(\eta) (-d\eta^2 + d\vec{x}^2) \,.
\ee


The electromagnetic field case is particularly simple: $A_\mu$ 
has zero conformal weight, that is it is unchanged 
under conformal trasformations. This is also true for 
$F_{\mu \nu}$ and for the action associated with it. 
For the case of fermions, one also has that the massless spin $1/2$ 
Lagrangian density is invariant under conformal trasformations 
with 
\be
\psi \rightarrow \tilde{\psi} = \Omega^{-3/2} \psi
\ee

Thus for the case of the electromagnetic field and massless 
fermions conformal invariance holds both for the action 
and the equations of motion.
Accordingly  
one then has that for a conformally flat metric the 
flat space-times results are reproduced and there is no particle 
production as the metric changes. 

For a massless conformally coupled scalar field, on the other hand, conformal 
invariance holds only for the equation of motion, while 
the action and its 
transformed version differ through the presence of a boundary (for 
the RW metric a total derivative) term. Boundary terms, 
which are classically associated with canonical transformations, do not
change
the equations of motion and conserved charges. However for a RW metric 
the matter Hamiltonian is not a conserved charge
and hence the time evolution is changed. 
Thus any attempt 
to quantize employing Lagrangians obtained on neglecting 
surface terms in RW metrics leads to questionable results, as also stated,
 but for a different reason, in \cite{fulling}.

\section{Rescaled scalar field}

Let us further discuss the consequences of the presence of a (time derivative) 
surface term in the scalar field action. As we have mentioned such a term
is classically associated with a canonical transformation and 
quantum mechanically will correspond to a unitary transformation (as 
is expected through the Poisson bracket - canonical commutator
correspondence, naturally on 
neglecting eventual anomalies). On expanding a massive non-minimally 
coupled scalar field in Fourier modes as in \cite{desitter}:
\be
\Phi = \frac{1}{\sqrt{V}} \sum_{\bf k}
\left[ e^{i {\bf k} \cdot {\bf x}} \Phi_{k} (\tau) + 
e^{- i {\bf k} \cdot {\bf x}} \Phi^*_{k} (\tau) \right]
\label{scalarFourier}
\ee
and on separating real and imaginary parts:
\be
\Phi_{k} (\tau) = \frac{1}{\sqrt{2}} \left( \phi_{k}^1 + i\phi_{k}^2
\right)
\ee
one obtains an action for each mode ${\bf k}, i$
\be
S_{\phi, k}^i = \frac{1}{2} 
\int a^3\,d\tau \left( \dot{\phi}_k^{i\,2} - \omega_k^2 \phi_k^{i\,2}
\right)
\,,
\label{scalaraction}
\ee
with $\omega_k^2 = \frac{k^2}{a^2} + \mu^2 +
\xi\,R$ and henceforth for semplicity we shall consider one mode $i, k$
and omit all such indices.
On rescaling $\phi = \zeta /a$ one immediately obtains the 
corresponding action:
\be
S_\zeta = \frac{1}{2} 
\int a \,d\tau \left[ \dot{\zeta}^2 - (\omega^2 - \frac{1}{6} R)
\zeta^2
- {1\over a}\frac{d}{d\tau} \left(\dot a \zeta^2 \right)\right]
\,.
\label{reaction}
\ee

 From the above actions, Eqs. (\ref{scalaraction}) and (\ref{reaction})
one obtains the
following Hamiltonians (related 
to evolution in proper time):
\begin{eqnarray}
H_\phi &=& \frac{1}{2a^3}
\pi_\phi^2 + \frac{a^3}{2} \omega^2 \phi^2 
\label{ham_phi} \\
H_\zeta &=& \frac{\pi_{\zeta}^2}{2 a} +
\frac{a}{2} \omega^2 \zeta^2 + \frac{\dot{a}}{2a} (
\pi_{\zeta} \zeta + \zeta \pi_{\zeta} ) 
\label{ham_zeta} \,,
\end{eqnarray}
where $\pi_\phi = a^3 \dot \phi$, $\pi_{\zeta} = a \left(
\dot{\zeta} - \frac{\dot a}{a}
\zeta \right) = \pi_\phi/a$ , we have used $R=6 \left( {\ddot a \over a}
+ {{\dot a}^2 \over {a}^2} \right) $ and one has 
\be
H_\zeta = H_\phi + \frac{\dot a}{2a} (\phi \pi_{\phi} +
\pi_{\phi} \phi) \,.
\ee 
It is easy to verify that the trasformation from $\phi$ to $\zeta$
is canonical:
\be
[\phi, \pi_\phi]_{PB} = [\zeta, \pi_\zeta]_{PB} = 1
\ee
and one relates the Poisson brackets (PB) to commutators in order to
canonically quantize the two systems. For the system with Hamiltonian
(\ref{ham_phi}) one can introduce the following operator, which
factorizes the Hamiltonian,
\be
\hat a_\phi = \left( \frac{a^3 \omega}{2 \hbar} \right)^{\frac{1}{2}}
\left( \hat \phi + i\,{\hat \pi_\phi \over a^3 \omega} \right) \,,
\ee
and the operator related to the linear invariant
\be
\hat b_\phi = \frac{1}{\sqrt{2 \hbar}} 
\left[{\hat \phi \over \rho} +
i\,\left(\rho \,\hat \pi_\phi - a^3 \dot{\rho} \hat \phi \right)\right] \,,
\ee
Similarly one may introduce the corresponding operators for the
Hamiltonian
(\ref{ham_zeta})
\begin{eqnarray}
\hat a_\zeta &=& \left( \frac{a \omega_D}{2 \hbar} \right)^{\frac{1}{2}}
\left( \hat \zeta + i\,{\hat \pi_\zeta \over a \omega_D} \right) \,,
\label{aoperator}
\\
\hat b_\zeta &=& \frac{1}{\sqrt{2 \hbar}} 
\left\{ {\hat \zeta \over a \rho} +
i\,\left[ a \rho (\hat \pi_\zeta + \dot a \hat \zeta) 
- (a \dot{\rho} + \dot a \rho) a \hat \zeta \right] \right\}\,,
\label{boperator}
\end{eqnarray}
where $\omega_D= (\omega^2 - \dot a^2/a^2)^{\frac{1}{2}}$ and in both
cases $\rho$ satisfies
\be
\ddot{\rho} + 3 \frac{\dot a}{a} \dot{\rho}
+ \omega^2 \rho = \frac{1}{a^6 \rho^3}
\ee
Of course the above operators satisfy 
$[\hat a,\hat a^\dagger]=[\hat b,\hat b^\dagger]=1$ and
remarkably $\hat b_\phi$ and $\hat b_\zeta$ {\em coincide} implying that
they lead to the same vacuum and Fock space. Actually this
should not be surprising since the classical solutions to the
equation of motion for $\phi$ and $\zeta/a$ are the same.

On using the above operators one obtains:
\begin{eqnarray}
\hat I &=& \hbar (\hat b_\phi^\dagger \hat b_\phi + \frac{1}{2} ) = 
\hbar (\hat b_\zeta^\dagger \hat b_\zeta + \frac{1}{2} )    \\
\hat H_\phi &=& \hbar \omega ( \hat a_\phi^\dagger \hat a_\phi + \frac{1}{2}) \\
\hat H_\zeta &=& \hbar \omega_D 
( \hat a_\zeta^\dagger \hat a_\zeta + \frac{1}{2} ) \label{accazeta}        
\end{eqnarray}
where $\hat I$ is an {\em invariant} and has time-independent eigenvalues,
while $\hat H_\phi$ and $\hat H_\zeta$ obviously do not and correspond to
the number of 
$\phi$ and $\zeta$ quanta times their respective energies.
It is clear from Eq. (\ref{scalaraction}) that it is only the $\phi$
quanta
that have a
particle interpretation and the $\phi$ and $\zeta$ quanta are related through
a Bogoliubov transformation
\be
\hat a_\phi =  \frac{1}{2} \hat a_\zeta [ \frac{\omega^{1/2}}{\omega_D^{1/2}}
+ \frac{\omega_D^{1/2}}{\omega^{1/2}} ] + \frac{1}{2} \hat a_\zeta^\dagger
[\frac{\omega^{1/2}}{\omega_D^{1/2}}
- \frac{\omega_D^{1/2}}{\omega^{1/2}} ] \label{Bogtrans}
\ee
corresponding to a squeezing \cite{schumaker}. Naturally the $\hat b$
and $\hat a$ are also related through a Bogoliubov transformation
and it is the $\hat b$ Fock space that describes the correct evolution.

Let us end by commenting that if we had omitted the surface term
in Eq. (\ref{reaction}) we would have obtained a third different
Hamiltonian (which, apart from overall scale factors, coincides with the 00
component of the scalar field energy-momentum tensor appearing in the
Einstein equations for $\xi =1/6 $) 

\be
\bar{H}_\zeta = \frac{1}{2 a} \bar{\pi}^2_\zeta + \frac{a}{2} ( \omega^2
- \frac{R}{6}) \zeta^2
\label{ham_skaz}
\ee
where $\bar{\pi}_\zeta = a \dot \zeta = \pi_\zeta + \dot a \zeta $.
Again one may study the quantum system and introduce both linear invariant
operators and operators factorizing the Hamiltonian. It is straightforward
to verify that the former agree with Eq. (\ref{boperator}), that is one obtains 
the same invariant vacuum and the Fock space. This is not surprising since the 
equations of motion again are unchanged. For the quantum Hamiltonian,
on the other hand, one obtains:
\be
\hat {\bar H}_{\zeta} = \hbar \bar \omega_D \left( \hat {\bar 
a}_{\zeta}^{\dagger} \hat {\bar a}_{\zeta} +{1\over 2} \right)
\ee
where $\bar {\omega}_D = ( \omega^2 - R/6 )^{1/2}$ and
\be
   \hat {\bar a}_{\zeta}^{\dagger} = {\left( {a\bar \omega_D 
   \over{2 \hbar}}\right)}^{1/2} \left[ \hat \zeta +{i\over{a\bar \omega _D}}
   \hat{\bar \pi}_{\zeta} \right]
\ee
which are related to the corresponding operators in Eq.(\ref{aoperator}) 
through a Bogolubov transformation:
\be
\hat {\bar a}_{\zeta}= {1\over 2} \hat a_{\zeta} \left[ \left( 1+i{\dot a \over 
{a \bar \omega_D}} \right) \left( {\bar \omega_D \over \omega_D} 
\right)^{1\over 2}
 + \left( {\omega_D \over \bar \omega_D} \right)^{1\over 2} \right] +
 {1\over 2} \hat a_{\zeta} ^{\dagger} 
 \left[ \left( 1+i {\dot a \over 
{a \bar \omega_D}} \right) \left( {\bar \omega_D \over \omega_D}
 \right)^{1\over 2}-\left( {\omega_D \over \bar \omega_D} \right)^{1\over 2}
  \right] 
\ee
It is important to note the different spectrum obtained with respect to
Eqs. (\ref{accazeta}) and (\ref{Bogtrans}). Thus rescaling and omitting 
the surface term  obtained will modify the quantum system and the spectrum of
the Hamiltonian, leading to questionable results for some physical 
quantities, while maintaining the invariant vacuum and Fock space which 
corresponds to leaving the equations of motion unchanged.
For example we may 
 consider the invariant vacuum ($ \mid 0 \rangle _b$) expectation values of the 
 quantum Hamiltonians given by eqs.($70$) and eqs.($74$) for $\xi =0$ (thus we 
 do not have the presence of derivatives of $a$ in the matter Hamiltonian
- we return to this at the end of the next section) in a de Sitter space and
consider the 
 limit as $ a \longrightarrow \infty $. One obtains \cite{desitter}:
 $$
{}_b \langle 0|\hat{H}_{\phi}|0 \rangle_b \simeq \frac{\hbar}{4} B \left(
\frac{\g}{\pi}
\right)^2 \quadcap
 a^{2 \bar{\nu}} \left[ \mu ^2 + H^2 \left( \bar{\nu} - \frac{3}{2}
\right)^2 \right]
 $$
 $$
{}_b \langle 0|\hat{\bar{H}}_{\zeta}|0 \rangle _b \simeq \frac{\hbar}{4} B
\left(
\frac{\g}{\pi} \right)^2 \quadcap
a^{2\bar{\nu}} \left[ \mu ^2 -2H^2 + H^2 \left( \bar{\nu} - \frac{1}{2}
\right)^2 \right]
 $$
 where $B$ is a constant and $\bar{\nu}^2= 9/4 - \mu ^2/ H^2 $. We
immediately
 note that the two expressions differ and in particular the first one vanishes
 for $\mu =0$ (of course non-leading terms remain). Similar considerations
hold on considering expectation values 
 with respect to other states. A detailed discussion of the 'squeezing' effect
 of the surface term for the $\xi=\mu=0$ case has been previously done
 \cite{polarski}.
 
\section{Conclusions}

The method of invariants is a particularly suitable tool in order to
investigate quantum effects in time-dependent external fields. 
On using a conserved
operator, one can find an invariant vacuum and an invariant Fock space,
which are preserved during the time evolution. 
In this paper we have
used this tool
in order to investigate the relation between particle production 
and the conformal properties of fields.

Motivated by the non conformal invariance of the action for the 
scalar field case \cite{desitter} - shown in section 4 for a general
space-time -, we have analyzed the electromagnetic
and Dirac fields. 
For the electromagnetic and the massless spinor fields the action 
is conformally invariant, and we have verified that there is no particle 
production in the Hamiltonian (which is equal to the energy density in
these cases). For the massive spinor case there is a non-trivial particle 
production, but, because of the Pauli exclusion principle, the number of 
particles in a given mode can never exceed one. 

We have gone beyond our previous paper \cite{desitter} 
and investigated, for cosmological spacetimes, some effects due to
the presence of the 
surface term which violates the conformal invariance of the action in the 
scalar field case with the following results: 

1. We verified that the use of rescaled fields is classically a
canonical trasformation and a unitary trasformation quantum mechanically.
On retaining all terms in the action we used
the rescaled fields in order to canonically quantize the system. and
obtained the same invariant vacuum and the same invariant Fock space as  
resulted on quantizing the original (non-rescaled) fields.  

2. The fact that the energy density of a non-minimally coupled scalar
field (which is the source of the Einstein tensor) differs from the 
canonically obtained Hamiltonians of the field and of the rescaled field
(which generate the correct time evolution for the two systems) implies that
the usual statement of no particle production - on neglecting 
quantum anomalies - in the 
massless conformally coupled scalar field should only be applied 
to the source of Einstein tensor.Indeed as we pointed out in the previous
 section apart from overall scale factors the
 rescaled scalar field Hamiltonian, obtained on omitting surface terms,
 coincides with the $00$ component of the scalar field energy momentum
tensor appearing in the Einstein equations for $\xi = 1/6 $.

3. The neglect of the surface term in the action for the
rescaled fields classically leads to
a third and different Hamiltonian (\ref{ham_skaz}), and so a different
time evolution.   
On canonically quantizing this system, one however gets the
same invariant vacuum and Fock space corresponding to the fact that
the equations of motion are unchanged. 

Let us end by noting that in all the above we have employed the usual 
canonical formalism for the scalar field starting from the action and
obtaining the Hamiltonian describing time evolution. Time evolution
for matter can also be obtained from an initial action containing both
gravitation and matter leading to the Wheeler-De Witt equation which, of course,
does not contain time. On performing a Born-Oppenheimer decomposition
and considering the semiclassical limit for gravity one is led to the 
Schrodinger (or Schwinger-Tomonaga ) equation for the evolution of matter
\cite{LakeTahoe}.
We feel it would be interesting to obtain a similar approach for the case
of a non-minimally 
coupled scalar field. However there is an essential difficulty since a 
non-minimal coupling leads to terms containing derivatives of the metric 
appearing in the matter lagrangian. This will lead to momenta conjugate to the
metric appearing in the matter Hamiltonian in contrast with what is usually
assumed for a Born-Oppenheimer (or adiabatic) approach. Nonetheless we hope to
return to this.

\end{document}